\documentclass{PoS}
\usepackage{amsmath,amssymb}
\usepackage{graphics,graphicx}
\usepackage{mcite}

\newcommand{\tr}[0]{\text{tr}}

\def\Dslash{\mathchoice
    {D\hskip-0.62em\raise0.2ex\hbox{$\displaystyle/$}\hskip0.2em}%
    {D\hskip-0.62em\raise0.2ex\hbox{$\textstyle/$}\hskip0.2em}%
    {D\hskip-0.5em\raise0.15ex\hbox{$\scriptstyle/$}\hskip0.2em}%
    {D\hskip-0.5em\raise0.15ex\hbox{$\scriptscriptstyle/$}\hskip0.2em}}

\def\kslash{\mathchoice
    {k\hskip-0.5em\raise0.2ex\hbox{$\displaystyle/$}\hskip0.2em}%
    {k\hskip-0.52em\raise0.2ex\hbox{$\textstyle/$}\hskip0.2em}%
    {k\hskip-0.4em\raise0.15ex\hbox{$\scriptstyle/$}\hskip0.2em}%
    {k\hskip-0.4em\raise0.15ex\hbox{$\scriptscriptstyle/$}\hskip0.2em}}
\def\Aslash{\mathchoice
    {A\hskip-0.62em\raise0.2ex\hbox{$\displaystyle/$}\hskip0.2em}%
    {A\hskip-0.62em\raise0.2ex\hbox{$\textstyle/$}\hskip0.2em}%
    {A\hskip-0.5em\raise0.15ex\hbox{$\scriptstyle/$}\hskip0.2em}%
    {A\hskip-0.5em\raise0.15ex\hbox{$\scriptscriptstyle/$}\hskip0.2em}}

\def\jslash{\mathchoice
    {j\hskip-0.35em\raise0.2ex\hbox{$\displaystyle/$}\hskip0.2em}%
    {j\hskip-0.35em\raise0.2ex\hbox{$\textstyle/$}\hskip0.2em}%
    {j\hskip-0.25em\raise0.15ex\hbox{$\scriptstyle/$}\hskip0.2em}%
    {j\hskip-0.25em\raise0.15ex\hbox{$\scriptscriptstyle/$}\hskip0.2em}}

\def\pslash{\mathchoice
    {p\hskip-0.5em\raise0.2ex\hbox{$\displaystyle/$}\hskip0.2em}%
    {p\hskip-0.5em\raise0.2ex\hbox{$\textstyle/$}\hskip0.2em}%
    {p\hskip-0.4em\raise0.15ex\hbox{$\scriptstyle/$}\hskip0.2em}%
    {p\hskip-0.4em\raise0.15ex\hbox{$\scriptscriptstyle/$}\hskip0.2em}}

\def\qslash{\mathchoice
    {q\hskip-0.5em\raise0.2ex\hbox{$\displaystyle/$}\hskip0.2em}%
    {q\hskip-0.5em\raise0.2ex\hbox{$\textstyle/$}\hskip0.2em}%
    {q\hskip-0.4em\raise0.15ex\hbox{$\scriptstyle/$}\hskip0.2em}%
    {q\hskip-0.4em\raise0.15ex\hbox{$\scriptscriptstyle/$}\hskip0.2em}}

\def\partialslash{\mathchoice
    {\partial\hskip-0.62em\raise0.2ex\hbox{$\displaystyle/$}\hskip0.2em}%
    {\partial\hskip-0.62em\raise0.2ex\hbox{$\textstyle/$}\hskip0.2em}%
    {\partial\hskip-0.5em\raise0.15ex\hbox{$\scriptstyle/$}\hskip0.2em}%
    {\partial\hskip-0.5em\raise0.15ex\hbox{$\scriptscriptstyle/$}\hskip0.2em}}

\title{Confinement, the Abelian Decomposition, and the Contribution of Topology to the Static Quark Potential}

\ShortTitle{The Abelian Decomposition}

\author{\speaker{Nigel Cundy}\\
         Lattice Gauge Theory Research Center, FPRD, and CTP, Department of Physics \&
    Astronomy,\\ Seoul National University, Seoul, 151-747, South Korea\\
        E-mail: \email{ndcundy@gmx.com}}

\author{Yongmin Cho\\
        Administration Building 310-4, Konkuk University,
Seoul 143-701, Korea}

\author{Weonjong Lee\\
         Lattice Gauge Theory Research Center, FPRD, and CTP, Department of Physics \&
    Astronomy,\\ Seoul National University, Seoul, 151-747, South Korea}

\abstract{
In the past few years, we have presented a new way of considering quark confinement. Through a careful choice of a Cho-Duan-Ge Abelian Decomposition, we can construct the QCD Wilson Loop in terms of an Abelian restricted field. The relationship between the QCD and restricted string tensions is exact; and we do not need to gauge fix, apply any path ordering of gauge links, or additional path integrals. This hints at why mesons are colour neutral.

Furthermore, the Abelian restricted field contains two parts: a Maxwell term, and a topological term. The topological term can describe magnetic monopoles and other topological objects, which can be studied both numerically and theoretically. By examining the topological part of the restricted field strength we have found evidence suggesting that these objects, which will contribute to confinement if present, are indeed there.

Previous studies have used simplifications, breaking the exact relationship between the restricted and QCD string tensions, but it was found that the topological term dominated the restricted string tension. Here we remove those simplifications, and show that the Abelian restricted field does indeed fully explain confinement. However, our results for how much of the restricted string tension arises from the topological objects show strong dependence on the lattice spacing and level of smearing, so we are not yet able to draw a definitive conclusion. 
}
\FullConference{The 32nd International Symposium on Lattice Field Theory,\\
		23-28 June, 2014\\
		Columbia University New York, NY}

\begin{document}

\section{Introduction}

   Our previous studies~\cite{Cundy:2012ee,*Cundy:2013xsa,*Cundy:2013pfa}, using the gauge invariant CDG (Cho-Duan-Ge) Abelian decomposition~\cite{Cho:1980,*Cho:1981,*Duan:1979,*F-N:98,*Shabanov:1999} to study quark confinement, have shown that, by judicious choice of the colour field $\theta$ that defines the decomposition, one can calculate the static potential of a Yang Mills gauge theory entirely in terms of a `restricted' Abelian field, which (unlike QED) may contain topological objects such as magnetic monopoles. The Abelian restricted field is far easier to analyse than full QCD. Abelian dominance is natural because it explains why only the Abelian gluons are related to confinement (i.e. hadrons are colour neutral).  Eventually, we intend to combine our lattice simulations with algebraic and numerical estimates of the free energy (following ~\cite{Cho:2012pq}) to show that these topological objects are preferred below the de-confinement transition. Here we concentrate on using numerical simulations to try to understand what causes confinement in the restricted theory. Another research group is using the CDG decomposition to study confinement in lattice gauge theories, but uses a different choice of decomposition~\cite{Kondo:2005eq,*Kondo:2010pt,*Shibata:2007pi}.

  The Abelian field contains two parts -- the Maxwell term (similar to QED), which is directly dependent on the QCD gauge field $A_\mu$, and the topological term (absent in QED), which only depends on $A_\mu$ indirectly through $\theta$. A common understanding is that this $\theta$ field can describe topological structures which might explain confinement. In analogy to QED, The Maxwell term is not expected to contribute to confinement. Previous studies have tried to isolate the topological part of the field and show its dominance (see, for example, \cite{Kondo:2005eq,*Kondo:2010pt,*Shibata:2007pi,Kronfeld1987516,*Suzuki:1989gp}). However, these studies have all had the weakness that the link between the restricted string tension and the QCD string tension is not exact. They have shown, in other words, that the restricted string tension is dominated by the topological part, but it is unclear how much bearing this has on the problem of full QCD. On the other hand, as far as we know for the first time, we here use a method which has an exact algebraic identity between the restricted and QCD string tensions. The main purpose of this work is to investigate whether the expected topological objects indeed dominate the string tension.

 Section \ref{sec:2} outlines the theoretical basis of our work, section \ref{sec:3} presents our numerical results, and section \ref{sec:4} concludes.
  \section{The Abelian decomposion}\label{sec:2}
  The CDG Abelian decomposition extracts an Abelian component from the gauge field.
  Unlike other approaches, our method requires no gauge fixing, nor arbitrary cuts of the field, nor additional path integrals.
  In the continuum, choose a field $\theta(x) \in SU(N)$ (we will explain how to select $\theta$ later). Construct a rotated basis $n_a = \theta \lambda_a \theta^\dagger$, with $\lambda_a$ a Gell-Mann matrix. Pick out the Abelian directions $n_j\equiv n_3,n_8,\ldots$. The decomposed fields $\hat{A}_\mu$ and $X_\mu$ satisfy
\begin{align}
A_\mu =& \hat{A}_\mu + X_\mu &
D_\mu[\hat{A}]n_j =& 0& \tr (n_j X_\mu) =& 0.
\end{align}
$\hat{A}$ is known as the restricted field. There is a known solution~\cite{Cho:1980,*Cho:1981,*Duan:1979,*F-N:98,*Shabanov:1999},
\begin{align}
\hat{A}_\mu =& \frac{1}{2} n_j \tr(n_j A_{\mu}) +\frac{i}{4g}[n_j,\partial_\mu n_j]; \nonumber\\
F_{\mu\nu}[\hat{A}] = &\frac{n_j}{2} \big[\partial_\mu \tr (n_j A_\nu) -\partial_\nu \tr (n_j A_\mu)\big] + \frac{i}{8g}n_j \tr(n_j[\partial_\mu n_k,\partial_\nu n_k]).\label{eq:FandA}
\end{align}
 The generalisation to the lattice is straight-forward. Starting with the gauge link $U_{\mu,x}$, we write
\begin{align}
&U_{\mu,x} = \hat{X}_{\mu,x} \hat{U}_{\mu,x} ,&U,\hat{U},\hat{X}\in {\rm SU}(N)&\nonumber\\
&\hat{U}_{\mu,x} n_{j,x+\hat{\mu}} \hat{U}_{\mu,x}^\dagger - n_{j,x} = 0 & \tr (n_{j,x}(\hat{X}_{\mu,x} - \hat{X}^\dagger_{\mu,x})) = 0&.\label{eq:latdefeq}
\end{align}
If there are multiple solutions, we select the one with maximal $\tr\; \hat{X}_\mu$.
 $\hat{U}$ corresponds to a gauge link constructed from $\hat{A}$, and
 $\hat{X}$ is related to $X$.
  If $\theta$ transforms under a gauge transformation as 
\begin{align}
U_{\mu,x} \rightarrow & \Lambda_x U_{\mu,x} \Lambda^\dagger_{x+\hat{\mu}} & \theta_x \rightarrow & \Lambda_x\theta_x,\label{eq:gttheta}
\end{align}
with $\Lambda_x \in {\rm SU}(N)$, then it is easy to find the transformations of $\hat{U}$ and $\hat{X}$
\begin{align}
\hat{U}_{\mu,x} \rightarrow & \Lambda_{x} \hat{U}_{\mu,x} \Lambda^\dagger_{x+\hat{\mu}} &\hat{X}_{\mu,x} \rightarrow &\Lambda_{x} \hat{X}_{\mu,x} \Lambda^\dagger_{x}.\label{eq:gthatu}
\end{align}
 Paths of gauge links constructed from $\hat{U}$ are gauge covariant: we do not need to gauge fix.

 Our aim is to extract the static potential from the Wilson Loop. We choose $\theta$ so the Wilson Loop for the restricted and QCD gauge fields are identical: for every link contributing to the Wilson Loop $\hat{U}_\mu \equiv {U}_\mu$. We then extend $\theta$ across all space by considering nested and stacked sets of Wilson Loops. This choice of $\theta_x$ contains the eigenvectors of the Wilson Loop starting and ending at position $x$. It is unique up to a $(U(1))^{N-1}$ transformation (which leaves $n_j$ invariant) and the ordering of the eigenvectors. Note that this requires calculating a new $\theta$ field for each Wilson Loop studied. The observed topological objects will depend on the Wilson Loop.
From (\ref{eq:latdefeq}), (\ref {eq:gttheta}) and (\ref{eq:gthatu}), $\theta^\dagger_x \hat{U}_{\mu,x} \theta_{x+\hat{\mu}}$ is Abelian and gauge invariant, so there is no need for path ordering or gauge fixing. The coloured field $X$ does not contribute to confinement, so mesons are colour-neutral.

 Both $F_{\mu\nu}[\hat{A}]$ and $\hat{A}_\mu$, defined in the continuum in equation (\ref{eq:FandA}), depend on two terms: (1) a function of both $A_\mu$ and $\theta$ (the Maxwell term); and (2) a function of $\theta$ alone (the topological term). We will denote the topological part of $F_{\mu\nu}[\hat{A}]$ as $H^{j}_{\mu\nu}\equiv \frac{i}{8g} \tr(n_j[\partial_\mu n_k,\partial_\nu n_k])$.

 Any SU(2) matrix (the theory for SU(3) proceeds in a similar way, but is too cumbersome to describe here) may be parametrised in terms of three variables $0\le a\le \frac{\pi}{2},c\in \mathbb{R},d \in \mathbb{Re}$
\begin{align}
\theta =& \left(\begin{array}{cc} \cos a & i \sin a e^{ic}\\
i \sin a e^{-ic}& \cos a\end{array}\right)\left(\begin{array}{cc} e^{id} & 0\\0& e^{-id}\end{array}\right),& \bar{\phi} =& \left(\begin{array}{cc} 0&i e^{ic}\\-ie^{-ic}&0\end{array}\right).\nonumber
\end{align}
 $d$ makes no contribution to $n_3 = \theta \lambda_3 \theta^\dagger$, so we can safely fix it to zero. The topological parts of $\hat{A}_\mu$ and $F_{\mu\nu}[\hat{A}]$ are
 ${\theta^\dagger\partial_\mu\theta} = \lambda_3{\sin^2a \partial_\mu c} + \bar{\phi}\cos2a \partial_\mu a $ and
 ${\tr(n_3[\partial_\mu n_3,\partial_\nu n_3])} = \partial_\mu a \partial_\nu c - \partial_\nu a \partial_\mu c$.

Selecting our Wilson loop in the $xt$ plane, we seek to map $a$ and $c$ to four dimensional Euclidean space. $c$ is undefined at $a = 0$ and $a = \pi/2$. Given that these are the maximum and minimum values of $a$, the singularities should occur at points rather than lines or surfaces, suggesting that it is advantageous to use a polar parametrisation of space-time. Keeping the rotational symmetry in the $yz$ plane manifest leaves two natural ways to parametrise the coordinates,
 $(t,x,y,z) = r(\cos \psi_3, \sin\psi_3 \cos\psi_2,\sin\psi_3\sin\psi_2\cos\psi_1,\sin\psi_3\sin\psi_2\sin\psi_1)$ and the system obtained through $t \leftrightarrow x$. The object is centred at $r=0$.
 There are two types of topological object available: the Wang-Yu magnetic monopole (e.g. $a = \psi_1/2, c = {\nu_{WY}} \psi_2$); and an object with $\pi_1$ topology (e.g. $a(r,\psi_1,\psi_2,\psi_3), c = {\nu_{T}}\psi_3$). {$\nu_{WY}$} and {$\nu_T$}, invariant under continuous gauge transformations, are integer winding numbers.
   Applying Stokes' theorem to the Abelian representation of the Wilson Loop gives a surface integral over the continuous part of $F_{\mu\nu}[\hat{A}]$ bound by line integrals around each topological singularity. These line integrals ($\sim\oint dx_\mu (\sin^2 a) \partial_\mu c$) are proportional to the winding number $\nu_T$ (the monopoles do not directly contribute). Since the number of these objects is proportional to the area of the Wilson loop, this leads to a linear static potential~\cite{cundyforthcoming}.
 
 One can calculate the field strength surrounding each of these topological objects. In analogy to electromagnetism, we characterise $H^{j}_{\mu\nu}$ in terms of `Electric' and `Magnetic' fields $\mathbf{E}^{j}$ and $\mathbf{B}^{j}$. The monopoles, with no contribution to the $E_x$ field, have either 1-dimensional lines of high field strength either in $B_x$, $B_y$ and $B_z$ parallel to the $T$ axis or in $B_x$, $E_y$ and $E_z$ parallel to the $X$ axis, depending on the choice of coordinate parametrisation. The $\pi_1$ objects, on the other hand, provide us with point-like structures in the $E_x$ field, accompanied by either
1-dimensional lines in  $B_x$, $B_y$ and $B_z$ parallel to the $T$ axis; or 
lines in  $B_x$, $E_y$ and $E_z$ parallel to the $X$ axis. The two sets of structures can only be distinguished by their contributions to the $E_x$ field. We expect to see $F_{\mu\nu}^j[\hat{A}]$ and $H_{\mu\nu}^j$ dominated by these two types of objects. 

\section{Numerical results}\label{sec:3}
\begin{figure}
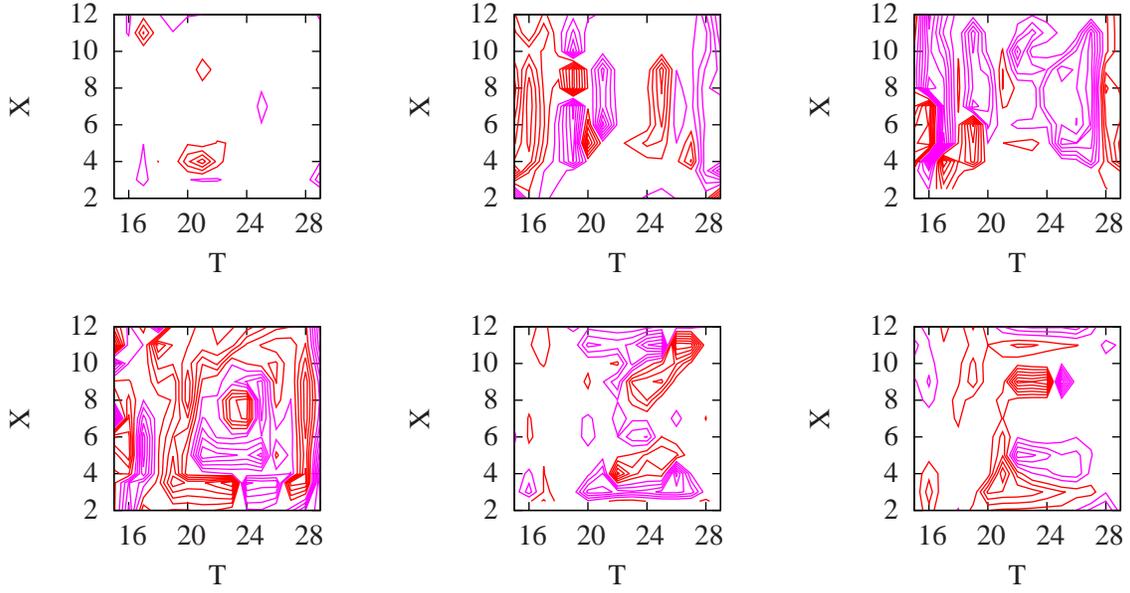

\vspace{-\baselineskip}
\begin{center}
{
 \begin{tabular}{ccc}
  \input{Fmunu_E_x_b8-52_70_xt.tex}&  \input{Fmunu_E_y_b8-52_70_xt.tex}&  \input{Fmunu_E_z_b8-52_70_xt.tex}\\
  \input{Fmunu_B_x_b8-52_70_xt.tex}&  \input{Fmunu_B_y_b8-52_70_xt.tex}&  \input{Fmunu_B_z_b8-52_70_xt.tex}
 \end{tabular}
}
\end{center}
\vspace{-.8cm}
\caption{The $X$ (left) $Y$ (middle) and $Z$ (right) components of the Electric (top) and Magnetic (bottom) Abelian Field Strengths. The contour plots show the field strength $H_{\mu\nu}^8$ (red contour lines positive field strength, purple lines negative field strength) on a slice of the lattice in the $XT$ plane (10 smearing sweeps).}\label{fig:1}
\vspace{-\baselineskip}
\end{figure}
Our numerical simulations are carried out with a tadpole improved Luscher-Weisz gauge action~\cite{TILW,*TILW2,*TILW4} in SU(3) quenched QCD at $\beta = 8.0,8.3$ and $8.52$ and lattice spacings (measured from the string tension) of around $0.14~\rm{fm}$, $0.11~\rm{fm}$, and $0.095~\rm{fm}$, with three ensembles on $16^3\times 32$ lattices and the $\beta 8.3L$ ensemble on a $20^3\times 40$ lattice. The plots show data from the $\beta =8.52$ ensemble; the other ensembles show similar results. There is no noticeable difference between the field strengths derived from $\lambda^3$ or $\lambda^8$. We applied stout smearing~\cite{Morningstar:2003gk,*Moran:2008ra} at parameters $\rho = 0.015$, $\epsilon = 0.0$ to smooth the gauge field and remove dislocations before computing $\theta$ or the Wilson Loop. The results discussed here were based on either ten or sixteen smearing steps. We are currently investigating different levels of smearing.

\begin{figure}
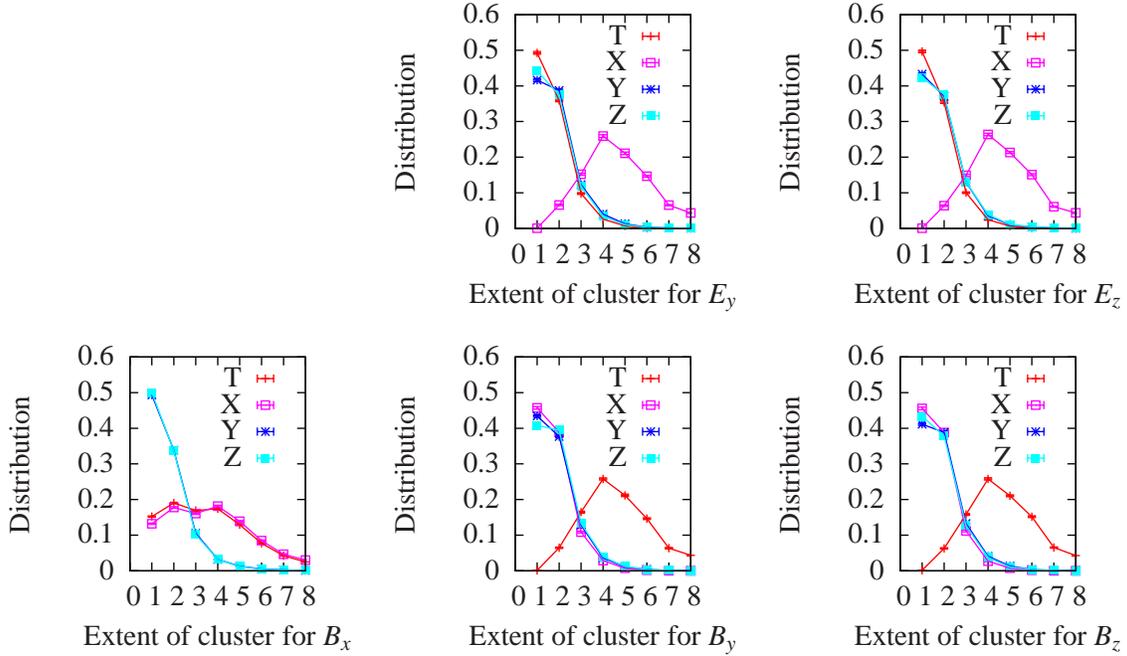

\vspace{-\baselineskip}
\begin{center}
{
 \begin{tabular}{ccc}
  &  \input{Fmunu_Cluster_extent_E_y_b8_52.tex}&  \input{Fmunu_Cluster_extent_E_z_b8_52.tex}\\
  \input{Fmunu_Cluster_extent_B_x_b8_52.tex}&  \input{Fmunu_Cluster_extent_B_y_b8_52.tex}&  \input{Fmunu_Cluster_extent_B_z_b8_52.tex}
 \end{tabular}
}
\end{center}
\vspace{-.8cm}
\caption{The extent in each spatial direction of structures of connected high field strength for the $H^8_{\mu\nu}$ field.}\label{fig:2}
\vspace{-\baselineskip}
\end{figure}

Figure \ref{fig:1} shows a contour plot for the components of $H^8_{\mu\nu}$, displaying the size and shape of the structures. In accordance with our model, we see point like objects in the $E_x$ component of the field, and lines along either the $t$ or $x$ axis (or both) in the other components of the fields. We use a cluster analysis to average geometric features across the whole ensemble. In figure \ref{fig:2}, we find clusters of connected sign-coherent high field strength, and measure how far these clusters extend in each direction. The structures of high field strength are indeed orientated exclusively in the predicted directions. We have also investigated the distribution of the number of nearest neighbours for each lattice site within the cluster -- these objects are one dimensional -- and whether the peaks in the $E_x$ field are correlated with high field strengths in the other fields -- they are~\cite{cundyforthcoming}. Thus the patterns in $H_{\mu\nu}$ are consistent with a liquid of Wang-Yu monopoles and the $\pi_1$ topological objects.


\begin{figure}
\vspace{-\baselineskip}
 \begin{center}
{
 \input{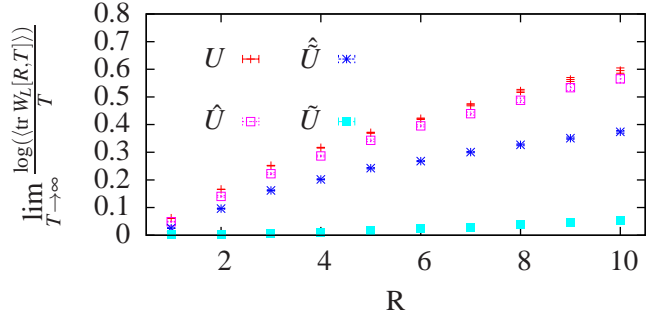}
}
\end{center}
\vspace{-.8cm}
\caption{The static quark potential calculated using a $T\times R$ Wilson Loop $W_L$ and the original ($U$), restricted ($\hat{U}$), topological ($\hat{\tilde{U}}$) and smeared $(\tilde{U})$ gauge fields (10 smearing sweeps).}\label{fig:st}
\end{figure}

   Neither the Maxwell nor topological parts are by themselves gauge invariant, but only their sum. The winding numbers of the topological objects within $\theta$ are invariant under continuous (differentiable with respect to the space/time coordinates) gauge transformations. However, `continuous gauge transformations' are problematic on the lattice. Thus any lattice study which directly separates the topological from the Maxwell part of the restricted gauge field will be gauge-dependent. Our alternative approach is described below. A preliminary investigation, which, like the other studies of the topological dominance of the restricted field strength, broke the identity between the restricted and original string tensions, confirmed the topological dominance found by other methods~\cite{Cundy:2012ee,*Cundy:2013xsa,*Cundy:2013pfa}. To measure the topological part of the string tension, we use the field
\begin{gather}
\hat{\tilde{A}}_{s,\mu} = \frac{n_j}{2} \tr(n_j \tilde{A}_{s,\mu}) + \frac{i}{4g} [n_j,\partial_\mu n_j]
\end{gather}
 We have replaced $A$ with a highly smoothed stout-smeared~\cite{Morningstar:2003gk,*Moran:2008ra} gauge field $\tilde{A}_s$ (calculated after $s$ additional smearing sweeps) while using the $\theta$ field derived from the original $A$. Enough smearing should remove any contribution from $\tilde{A}_s$, so only contributions from $\theta$ and a gauge transformation remain. 
 However, the string tension $\rho_{\theta,\tilde{A}_s}$ calculated from $ \hat{\tilde{A}}_{s,\mu}$ depended on $s$ even after 2500 smearing sweeps. However, we found that for $s$ between $600$ and $2500$, within our statistical errors $\rho_{\theta,\tilde{A}_s}  = \text{constant} + \rho_{\tilde{A}_s}$, with $\rho_{\tilde{A}_s}$ the string tension calculated from the smeared $\tilde{A}_s$ field, allowing an extrapolation to the point where the effects of $\tilde{A}$ are negligible. The string tension is shown in figure \ref{fig:st} and table \ref{tab:1}. The bottom row of the table shows the ratio of the topological component of the string tension and the full string tension.
\begin{table}
 \begin{center}
\begin{tabular}{|c|cccc|| ccc|}
\hline $\beta$&$8.0$&$8.3$&$8.52$&$8.3L$&$8.0$&$8.3$&$8.52$
 \\
 \hline
  $U$&0.098(2)&0.060(1)&0.0420(8)&0.0598(8)&0.095(4)&0.074(2)&0.049(1)
  \\
 $\hat{U}$&0.092(4)&0.058(2)&0.042(2)&0.063(3)&0.105(5)&0.079(2)&0.051(2)
  \\\hline
  $\hat{\tilde{U}}_{2500}$&0.038(1)&0.0256(9)&0.0221(7)&0.028(1)&0.038(2)&0.0335(10)&0.031(1)
  \\\hline
$\frac{\hat{\tilde{U}}_{2500} - \tilde{U}_{2500}}{U}$&0.27(1)&0.28(1)&0.32(1)&0.32(2)&0.25(2)&0.31(2)&0.44(3)

\\\hline
  \end{tabular}
    \end{center}
\vspace{-.2cm}
    \caption{Our fits for the string tension after 10 (left) and 16 (right) initial smearing steps.}\label{tab:1}
    \vspace{-\baselineskip}
\end{table}
The static potential for the restricted and original gauge fields are in good agreement (the difference is caused by different samples of Wilson Loops used in the two measurements). However, while the topological field contributes to the overall string tension, the contribution is smaller than expected, though it increases as we apply more smearing. At larger numbers of smearing sweeps, we see a considerable increase in the topological contribution to the string tension as the lattice spacing decreases, so it is possible that this small value is a lattice artefact combined with distortions from unphysical dislocations. Since the value of the QCD Wilson Loop ($U$) still increases with smearing, more smears might be needed to remove all the dislocations. To say anything definitive, we need to complete our study of the effects of smearing on this quantity, and then repeat the calculation at finer lattice spacings.
\section{Conclusions}\label{sec:4}
 We have shown that the string tension for the Abelian restricted field is identical to the string tension of the actual Yang-Mills gauge field, and therefore a proof of confinement of the (easier to analyse) restricted field will equally demonstrate confinement in full Yang-Mills theory. This is alone enough to demonstrate that the mesons at least are colour neutral. We have identified topological structures that can appear in the topological part of the restricted field: the familiar $\pi_2$ Wang-Yu magnetic monopoles, and other $\pi_1$ objects. These $\pi_1$ objects drive confinement. We do not make use of monopole condensation and the dual-Meissner effect: our proposed confinement mechanism seems to be something different. The structures observed in the field strength are consistent with expectations from our model. However, our analysis of whether the topological term is the dominant contribution to the restricted potential is not yet conclusive, since there are ambiguities related to the amount of smearing and there seem to be some lattice artefacts. To resolve these ambiguities requires a more detailed study. This is surprising, since earlier results (including our own using the same method to calculate the topological contribution) where the link between the restricted and full string tensions was only approximate lacked this ambiguity. Questions have also been raised as to whether our method of isolating the topological term by applying the Abelian decomposition on a highly smoothed field is valid~\cite{Cho:2010cw}, which might explain why we see a lower than expected topological contribution to the string tension. 

\section*{Acknowledgements}
Numerical calculations used servers at Seoul National University. The authors thank "BK21 Plus Frontier Physics Research Division, Department of Physics and Astronomy, Seoul National University, Seoul, South Korea" for financial support.
This research was supported by Basic Science Research Program through the National Research Foundation of Korea(NRF) funded by the Ministry of Education(2013057640).
The research of W. L. is supported by the Creative Research Initiatives Program(No. 2014-001852) of the NRF grant funded by the Korean government(MSIP).
 W. L. would like to acknowledge the support from KISTI supercomputing center through the strategic support program for the supercomputing application research(No. KSC-2013-G3-01).
 YMC is supported in part by NRF grant (2012-002-134)
funded by MEST.

 \bibliographystyle{JHEP_mcite.bst}

\bibliography{weyl}



\end{document}